%Paper: hep-th/9210041
%From: Cobi Sonnenschein <COBI@TAUNIVM.TAU.AC.IL>
%Date: Thu, 08 Oct 92 15:33:27 IST

\input phyzzx

%macropackage=phyzzx
%\baselineskip=10pt
\def\co{cohomology\ }
\def\G{${G\over G}\ $}

\def\F{{\cal F}(J,I)\ }
\def\c#1#2{\chi_{#1}^{#2}}
\def\p#1#2{\phi_{#1}^{#2}}
\def\b#1{\beta_{#1}}
\def\g#1{\gamma_{#1}}
\def\r#1#2{\rho_{#1}^{#2}}
\def\jt#1#2{{J^{(tot)}}_{#1}^{#2}}
\def\y{|phys>}
\def\t{\tilde}
 \def\pa{\partial}

\def \jg{J^{(gh)}}
\def \jb{J^{(BRST)}}
\def \pa{\partial}

\def\S#1{$SL({#1},R)$}
\def\A {$A_{N-1}^{(1)}$}

\def\W{$W_N\ $ }
\def\G{${G\over G}\ $}
\def\GH{${G\over H}\ $}
\def\tGH{twisted\ ${G\over H}\ $}

\def\TKS{twisted Kazama-Suzuki\ }
\def \vh{\vec{\t {\cal H}}}
 \def\t{\tilde}
 \def\pa{\partial}
\def\bpa{\bar\partial}
\def\Frs{{\cal F}_{r,s}\ }

%*********************************************
\def\cmp#1{{\it Comm. Math. Phys.} {\bf #1}}

\def\pl#1{{\it Phys. Lett.} {\bf #1B}}
\def\prl#1{{\it Phys. Rev. Lett.} {\bf #1}}
\def\prd#1{{\it Phys. Rev.} {\bf D#1}}

\def\np#1{{\it Nucl. Phys.} {\bf B#1}}

\def\jmath#1{{\it J. Math. Phys.} {\bf #1}}
\def\mpl#1{{\it Mod. Phys. Lett.}{\bf A#1}}

%%%%%%%%%%%%%%%%%%%%%%%%%%%%%%%%
\REF\Wym{E. Witten, \cmp {117} (1988) 353.}
\REF\dvv{R. Dijkgraaf, E. Verlinde, and H. Verlinde,
\np {352} (1991) 59;
``Notes On Topological String Theory And 2D Quantum Gravity,'' Princeton
preprint PUPT-1217 (1990).}
\REF\Novel  {  J. Sonnenschein and S.
Yankielowicz ,`` Novel Symmetries of Topological
Conformal Field Theory", TAUP- 1898-91 August  1991.}
\REF\Wcs{E. Witten \cmp {121} (1989) 351.}
\REF\MS{D. Montano, J. Sonnenschein ,  \np {324} (1989) 348,
J. Sonnenschein \prd {42} (1990) 2080.}
\REF\DDK{F. David \mpl {3}  (1988) 1651;\break J. Distler and H. Kawai \np
{321}
(1989) 509.}
 \REF\Pol{A. M. Polyakov, \mpl  {2}  (1987) 893.}
\REF\BKDSGM{D. Gross and A. A. Migdal \prl {64} (1990) 127;\break
M. Douglas and S. Shenker \np {335} (1990) 635;\break E. Brezin and V. A.
Kazakov, \pl {236} (1990) 144}
\REF\Konts{M. Kontsevich \cmp {147} (1992) 1.}
\REF\us  {O. Aharony,O. Ganor N. Sochen  J. Sonnenschein and S.
Yankielowicz ,``Physical states in the \G models and two dimensional
gravity", TAUP- 1947-92 April 1992.}
\REF\uss  { O. Aharony, J. Sonnenschein and S.
Yankielowicz ,``  \G models and \W strings" , TAUP- 1977-92 June 1992
(to be published in \pl {} ).}
\REF\usss  {O. Aharony,O. Ganor  J. Sonnenschein and S.
Yankielowicz ,``On the \tGH topological models", TAUP- 1990-92 August 1992.}
\REF\Wgg{E. Witten, ``On Holomorphic Factorization of WZW and Coset Models"
IASSNS-91-25.  }
\REF\SY{M. Spiegelglas and S. Yankielowicz `` \G Topological Field Theories by
Cosetting $G_k$ TAUP-1934
;``Fusion Rules As Amplitudes in $G/G$ Theories,'' Technion PH- 35-90}
\REF\Comment{ For a comment about a possible difficulty in case of non-compact
groups see ref. [\uss]}
\REF\Rab {K. Bardacki, E. Rabinovici, and B. Serin \np {299} (1988) 151.}
\REF\GK{
  K. Gawedzki and A. Kupianen , \pl {215} (1988) 119, \np
     {320} (1989)649.}
\REF\KS{D. Karabali and H. J. Schnitzer, \np {329} (1990) 625.}
\REF\KaSu{ Y. Kazama and H. Suzuki, \np {321} (1989).}
\REF\PW{A. Polyakov and P. B. Wigmann \pl {131} (1983)  121.}
\REF\BT{ R. Bott L. W. Tu
``Differential
Forms  in Algebraic Topology", Springer-Verlag  NY  1982.}
\REF\Wak{M. Wakimoto,   \cmp   {104}  (1989) 605.}

\REF\BMP{P. Bouwknegt, J. McCarthy and  K. Pilch Cern Preprint
TH-6162/91}
\REF\BMPN{P. Bouwknegt, J. McCarthy and  K. Pilch \pl {234} (1990) 297,
\cmp {131} (1990) 125.}
\REF\FeFr{B. Feigin and E. Frenkel \pl {246} (1990) 75.}
\REF\BLNW{ M. Bershadsky, W. Lerche, D. Nemeschansky and N. P. Warner
``A BRST operator for  non-critical W strings" CERN-TH  6582/92.}

\REF\KK{V. G. Kac and  D. A. Kazhdan   {\it Adv. Math } {\bf  34}  (1979) 79.}
\REF\BF{D. Bernard  and G. Felder \cmp {127}  (1991)  145.}

\REF\FaLu{ Fateev and Lukeanov
{\it Int. Jour. of Mod. Phys. }{\bf A31} (1988)
507.}

\REF\nota{For the definition of these fields see ref. [\uss]}
\REF\BerK{M. Bershadsky and I. Klebanov \np {360} (1991) 559.}
 \REF\LZ{B. Lian and G. Zuckerman \pl   {254}  (1991) 417.}
 \REF\Gepner{D. Gepner \cmp {141} (1990) 381.}

%%%%%%%%%%%%%%%%%%%%%%%%%%%%%%%%%%%%%%%%%%%%%%%%%%%%%
 %title page
\rightline{TAUP- 1999-92}
\date{Sept 1992}
\titlepage
\vskip 1cm
\title{Physical States in Topological Coset Models. }
\author { J. Sonnenschein
\footnote{*}{ Lecture delivered at the Nato workshop  on ``Low Dimensional
Topology and Quantum Field Theory" Sept. 1992, Isaac Newton Institute of
Mathematical Science.}  } \address{ School of Physics and Astronomy\break
Beverly
and Raymond Sackler \break Faculty of Exact Sciences\break
Ramat Aviv Tel-Aviv, 69978, Israel}
\abstract{ Recent results about topological coset models are summarized.
The action of a  topological \GH coset model ($rank\ H = rank\ G$) is
written down as a sum of ``decoupled" matter,
gauge and ghost sectors.  The
physical states are in the cohomology of a
BRST-like operator that relates
these secotrs.
The cohomology on a free field  Fock space  as well as on an irreducible
representation of the ``matter"
Kac-Moody algebra are extracted.  We compare the results
with  those of  $(p,q)$ minimal  models coupled to gravity
 and with $(p,q)$ $W_N$ strings for  the case of
$A_1^{(1)}$ at level $k={p\over q}-2$
and $A_1^{(N-1)}$ at level $k={p\over
q}-N$ respectively.}
\overfullrule=0pt
%**********************************************************
\singlespace
\section{ From TQFT's to topological coset models}
A Topological quantum field theory (TQFT)\refmark\Wym is a QFT in which all the
``observables", namely, all correlators of ``physical operators", are invariant
under any arbitrary deformation of
 $g_{\alpha\beta}$ the metric of the underlying space-time.
 Given a set of physcial operators $F_i[\Phi^a (x_i)]$ which are functionals
of  the  fields $\Phi^a (x)$, $a=1,...p$ of the theory and which are invariant
under the symmetries of the theory, then the theory is topological iff
$$\delta_{g_{\alpha\beta}}<\prod_i F_i[\Phi^a (x)]>=0.\eqn\mishtop$$
for any product which is a Lorentz scalar.
In particular this definition implies that all correlators are independent
of distances between the operators.
In a theory where  the energy-momentum  tensor
$T_{\alpha\beta}$ is exact under a BRST-like  symmetry operator $Q$, namely,
$T_{\alpha\beta}=\{Q, G_{\alpha\beta}\}$, and provided that all physical
operators  are in the  cohomology of $Q$  then for every product defined above
 property \mishtop\ is obeyed. Thus, it is a TQFT.
 Schematically the proof of the latter statement is the following
$$\eqalign{ &\delta_{g_{\alpha\beta}}<\prod_i F_i[\Phi^a (x_i)]>=
\delta_{g_{\alpha\beta}}[\int D\Phi^a e^{iS(\Phi^a)}\prod_i F_i[\Phi^a
(x_i)]\cr
&=\int D\Phi^a e^{iS(\Phi^a)}\int dx'T_{\alpha\beta} (x')\prod_i F_i[\Phi^a
(x_i)]\cr
& =\int D\Phi^a e^{iS(\Phi^a)}\int dx'\{Q, G_{\alpha\beta}\} (x')\prod_i
F_i[\Phi^a (x_i)]\
=<\{ Q,   \}>=0\cr} \eqn\mishind$$

A  two-dimensional theory  which is  a TQFT as well as conformal,
namely, $T_{zz}=T(z)$ and $ T_{\bar z\bar z}=\bar T(\bar z)$ and similarly for
all the symmetry currents, is a topological conformal field theory
(TCFT).
Any  theory which has the following algebra
 $$\eqalign{ T(z) =&\{ Q, G(z)\} \cr  \jb =&\{ Q ,
j^\#(z)\}\cr}
\eqn\mishbalgebra$$
where $Q=\oint \jb(z)$ is the BRST charge is a TCFT.
This algebra  is referred to as the   TCFT algebra\refmark\dvv and in fact it
is
a sub-algebra of a larger algebra which is shared by the TCFT's.\refmark\Novel
 An imediate consequence
of the fact that $T$ is exact is that the theory has a vanishing Virasoro
anomaly.

 Obviously every conformal field theroy coupled to 2-D gravity is a TCFT
after integration over all metric degrees of freedom.
We now define the concept of a ``topological model" which  is a
TQFT without the introduction and integration over the metric. In more than two
dimensions   examples of such models are the Chern-Simons theory and the four
dimensional theory which corresponds to  the Donaldson
invariants.\refmark\Wym In two dimensions an example is the theory of flat
gauge
connections.\refmark\MS We are now ready to introduce the  notion of
``topological coset models" which are  gagued WZW models that are also
topological models.

Gravitational models in two dimensions
were a subjecct of intensive study in recent years. Various different
approaches were invoked in this domain of research. Starting from continuum
Liouville theory and the light-cone world sheet formulation\refmark\Pol through
matrix\refmark{\BKDSGM} models,  KdV hierarchies to the Kontsevich
integral\refmark\Konts. Yet another possible approach is that of the
topological
coset models.  Here we summarize some recent work\refmark{\us,\uss,\usss} in
extracting the space of physical states of those models and reveal some
correspondence between these models and matter models
 like the $(p,q)$ minimal models coupled to two dimensional gravity

 \section{ The quantum action of a \tGH model}
 The  classical action  of the \tGH model\refmark\Wgg  is that of level $k$
twisted supersymmetric  $G$-WZW model coupled to gauge fields in the algebra of
$H\in G$.  In other words it is the usual \GH model with an extra  set of
$(1,0)$
anti-commuting ghosts where the dimension one fields take their values in the
positive roots of \GH and the dimension zero fields in the negative ones.
The action of the model reads
 $$ \eqalign{S_{(tKS)} &= S_k(g,A,\bar A) + S_{(gh)}^{G\over H}\cr
  S_k(g,A,\bar A)  &= S_{k}(g)
-{k\over 2\pi} \int_{\Sigma}d^2 z Tr_G[g^{-1}\pa g \bar A_{\bar z} +  g\bar \pa
g^{-1} A  - \bar A g^{-1}A  g  + A \bar A  ]\cr
S^{G\over H}_{gh}&={i\over {2\pi}}\int d^2z \sum_{\alpha\in{ G\over
H}}[\rho^{+\alpha}(\bar D \chi )^{-\alpha}+ \bar\rho^{+\alpha}( D
\bar\chi)^{-\alpha}]\cr}
\eqn\mishwzw$$
where $g\in G$ and $S_k(g)$ is the WZW action at level $k$.
In the case that $H=G$ the model coincides with the \G
model.\refmark\SY\  In the case that $\Sigma$ is topologically trivial the
gauge
fields
can be parametrized as follows $A=ih^{-1}\pa  h ,  \bar  A =i\bar h\bpa
\bar h^{-1}$ where $h(z), \bar h(z) \in H^c$.\refmark\Comment
 For the $U(1)$ parts of $H$
we take $A=i\pa {\cal H}_s$ where $s$ goes over the $U(1)$ factors.
The WZW part of the  action
then\refmark{\Rab,\GK,\KS}
takes the form
 $$S_k(g,A) =S_k(hg\bar h)
-S_k(h\bar h) \eqn\mishwzwh$$ The Jacobian  of the change of variables
introduces a   dimension $(1,0)$ system of anticommuting ghosts $\chi$ and
$\rho$ in the adjoint representation of $H$. The WZW  action thus
 becomes
$$S_k(g,A) =S_k(hg\bar h) -S_k(h\bar h) +{i\over {2\pi}}\int d^2z
Tr_H[\rho\bar D \chi + \bar\rho D \bar\chi] \eqn\mishwzwh$$ where
$D\chi=\pa\chi -i[A,\chi]$.  One then fixes the gauge by setting $\bar h=1$
which
implies $\bar A=0$ and redefining $hg\rightarrow g$. We shall treat in detail
the case
 of  $G=SU(N)$ and
$H=SU(N_1)\times ...\times SU(N_n)\times U(1)^r$ with $r =N-1-\sum_{I=1}^n
N_I+n$,  the gauge fields $A$ take the form $A= i\sum_{I=1}^n {h^{(I)}}^{(-1)}
\pa h^{(I)} +i\sum_{s=1}^r \pa {\cal H}_s$ and  the \TKS\refmark\KaSu action is
given by
 $$ \eqalign{S_{(tKS)} &= S_k(g) - \sum_{I=1}^n S_k(h^{(I)}) -
{k\over 4\pi  }\int d^2 z  \sum_{s=1}^r \pa {\cal H}_s\bpa {\cal H}_s\cr
  &+{i\over {2\pi}}\int d^2zTr_H[\rho\bar D \chi + \bar\rho \pa\bar\chi]
+{i\over {2\pi}}\int d^2z \sum_{\alpha\in{ G\over
H}}[\rho^{+\alpha}( \bpa \chi )^{-\alpha}+ \bar\rho^{+\alpha}( D
\bar\chi)^{-\alpha}]\cr}
\eqn\mishwzwHi$$
To achieve in the level of the action a complete decoupling of the  matter,
gauge and ghost sectors, one has to perform now a chiral rotation to eliminate
the coupling between the ghost and $A$ in  \mishwzwHi. This can be performed by
an explicit computation of the change in the measure of the ghosts or by
 non-abelian bosonization of the latter and using  the Polyakov Wiegmann
formula\refmark\PW
in the corresponding WZW actions. These two approaches lead to the same
result\refmark\usss  which for $rank\ H = rank\ G$
is  the following quantum action.

   $$ \eqalign{S_k =&S_k(g)
 + \sum_{I=1}^nS_{-(k +C_G +C_{H^{(I)}}) }(h^{(I)})+\cr
 &{1\over 2\pi }\int
d^2z[\sum_{s=1}^r \pa {\cal H}_s\bpa {\cal H}_s
 + i\sqrt{2\over
k+C_G}(\vec\rho_G-\vec\rho_H)\cdot \vh R] \cr
&+{i\over{2\pi}}\int d^2z Tr_H[\bar\rho \pa \chi + \rho\bar\pa\chi]
+{i\over{2\pi}}\int d^2z \sum_{\alpha\in{ G\over
H}}[\rho^{+\alpha}(\bar \pa \chi )^{-\alpha}+ \bar\rho^{+\alpha}( \pa
\bar\chi)^{-\alpha}] \cr}
 \eqn\mishwzwh$$
where we have  normalized the $\vh$ fields to be free  bosons, and
$\vec\rho_G$ and $\vec\rho_H$ are half the sums of the
positive roots of $G$ and $H$ respectively. The action is composed of three
decoupled sectors: the matter sector, the gauge sector and the ghost sector
involving  ghosts in $H$ and \GH.
 \section{ The algebraic structure of the \TKS model}
The next step in analyzing the \tGH models is to determine under what
conditions are they TCFT's. For that purpose one needs to derive the
  algebraic
structure of these models.
 We start with the Kac-Moody algebra associated with  the group $H$. We define
the currents ${\jt {} {a}}_I$ for each  non-abelian group factor  $H^{(I)}$,
and
for each $U(1)$ group, as
 $${\jt {} {a}}_I   =J^a +I^a  +if ^a_{bc}\r {} b \c {} c
+i\sum_{\alpha\beta\in {G\over H}} f^a_{\alpha,-\beta}\rho^\alpha\chi^{-\beta}.
\eqn\mishbJ$$  where $J^a,\ I^a$  are the contributions of the $g$ and $h$
sectors respectively.  The contribution of the   ghost currents
consists of both
the $H$ and \GH parts to be denoted by $J^a_H$ and $J^a_{G\over H}$
respectively.
  The level of these currents vanishes
 $$ k_I^{(tot)} = k
-(k+C_G+C_{H_I} ) + 2C_{H_I} +(C_G-C_{H_I}) =0. \eqn\mishbk$$ For the
abelian case there is a similar expression now with  $C_{H_I}=0$.

The energy momentum tensor $T$ can be decomposed, in a way which will be found
later to be natural,  into $T=T^H+T^{G\over H}$ as follows
$$ \eqalign{ T(z) &= {1\over2( k+c_G )} g_{\t a\t b}:J^{\t a} J^{\t b}: -
{1\over 2(k+c_G )}g_{ab}:I^a I^b: + g_{ab}\r  {} a \pa \c   {} b
\cr &-{\sqrt{2}\over k+C_G}(\vec\rho_G-\vec\rho_H)\pa\vec I
+ \sum_{\alpha\in{ G\over
H}}\rho^{+\alpha}( \pa \chi )^{-\alpha}\cr
T^H(z) &= {1\over2( k+c_G )} g_{ab}:(J^a +J^a_{G\over H})(J^b+J^b_{G\over
H}):-{1\over 2(k+c_G )}g_{ab}:I^a I^b:
\cr&-{\sqrt{2}\over k+C_G}(\vec\rho_G-\vec\rho_H)\cdot\pa(\vec J +\vec I +\vec
J_{G\over H})
 + g_{ab}\r  {} a \pa \c   {} b\cr
T^{G\over H}(z) &= {1\over 2(k+c_G )}g_{\t a\t b}:J^{\t a} J^{\t b}: -{1\over2(
k+c_G )} g_{ab}:(J^a +J^a_{G\over H})(J^b+J^b_{G\over H}): \cr
&+{\sqrt{2}\over
k+C_G}(\vec\rho_G-\vec\rho_H)\cdot\pa(\vec J  +\vec J_{G\over H})
+\sum_{\alpha\in{ G\over H}}\rho^{+\alpha}( \pa \chi )^{-\alpha}
 \cr}\eqn\mishbT$$
where $\t a$ and $\t b$ go over the adjoint of  $G$.
$\vec J$, $\vec I$ and $\vec J_{G\over H}$ are the Cartan-subalgebra currents
given in the basis in which $[J^i_n, J^j_m] = k n \delta ^{ij}\delta_{m+n}$.
The total Virasoro central  charge is found to be

$$ c= {kd_G\over k+C_G} +\sum_{I=1}^n {(k+C_G+C_{H^I})d_{H^I}\over k+C_G} +r
-2d_H -(d_G-d_H) +6[\sqrt{2\over k+C_G}(\vec\rho_G-\vec\rho_H)]^2 =
0\eqn\mishcz$$ where we have  used, assuming $G$ is a simply laced group,
 the relations  $12\rho_G^2 =d_GC_G$, and
$\vec\rho_H\cdot (\vec\rho_G-\vec\rho_H)=0$.

Upon gauge fixing, the gauge invariance is transformed into a BRST symmetry
generated by a dimension one current $\jb$. This current has a dimension two
partner $G$. These two anti-commuting currents are given by
$$ \eqalign{ \jb=&g_{ab}\c  {} a[J^b + I^b + J_{G\over H}^b + \half {\jg}^b_H]
\cr =&g_{ab}\c {} a[J^b + I^b + {i\over 2}f^b_{cd}\r {} c \c {} d  + {i}
f^{b}_{\gamma,-\beta}\rho^\gamma\chi^{-\beta}]\cr
G^H=&{g_{ab}\over 2(k +c_G)}\r  {} a [J^b - I^b +
if^{b}_{\gamma,-\beta}\rho^\gamma\chi^{-\beta}]
-{\sqrt{2}\over
k+C_G}(\vec\rho_G-\vec\rho_H)\cdot\pa\vec\rho
  \cr}\eqn\mishbGJ$$ It is
straightforward to realize that $T^H(z)$ is BRST exact
$$T^H(z)=\{Q^{(BRST)},G^H(z)\} \eqn\mishQGT$$ where $Q^{(BRST)}=\oint dz
J^{(BRST)}$. As was shown in ref.[\Wgg] the addition of the coset ghosts
turned the model into a twisted $N=2$ model. The twisted $N=2$ algebra is
generated
by $Q^{G\over H}$ and $G^{G\over H}$ given by
their $N=2$ counterparts\refmark\KaSu  with the fermions replaced with ghosts.
$$\eqalign{ Q^{G\over H}=
\sum_{\alpha\beta\gamma\in {G\over H}}\c {} {-\alpha}(J^\alpha + {i\over 2}
f^{\alpha}_{\gamma,-\beta}\rho^\gamma\chi^{-\beta})\cr
G^{G\over H}={1\over k+C_G}\sum_{\alpha\beta\gamma\in {G\over H}}\r {}
{\alpha}(J^{-\alpha} + {i\over 2}
f^{-\alpha}_{\gamma-\beta}\rho^\gamma\chi^{-\beta}).\cr }\eqn\mishQG$$
$T^{G\over H}$ defined above is exact with respect to $ Q^{G\over H}$
$$T^{G\over H}=\{Q^{G\over H},G^{G\over H}\}. \eqn\mishQGH$$
The various  $Q's$ and $G's$ obey the following anti-commutation relations:
$$ \eqalign{\{Q^{G\over H},Q^{(BRST)}\}&=
\{Q^{G\over H},G^H\}=\{Q^{(BRST)},G^{G\over H}\}= 0\cr
\{Q^{(BRST)},Q^{(BRST)}\}&= \{Q^{G\over H},Q^{G\over H}\}=\{G^{G\over
H},G^{G\over H}\}= 0\cr
\{G^{H},G^H\}&={1\over 4(k+C_G)^2} f_{abc}\r {} a  \r {} b \jt {} c
\cr}\eqn\mishGGQQ$$ The fact that $G^H$ is not nilpotent is shared by several
other TCFT's, in particular the \G models.
  Defining now the combined generators
$$ Q= Q^{(BRST)} +Q^{G\over H}\qquad  G= G^H + G^{G\over H} \eqn\mishQG$$
we find that (for $rank\  G =rank \ H$) the following relations hold
 $$\eqalign{ T(z) =&\{ Q, G(z)\} \cr  \jb =&\{ Q ,
j^\#(z)\}\cr
\jt {} a =&\{ Q ,\r {} a \}\cr}
\eqn\mishalgebra$$
where $\jt {} a$ denotes a current in the  algebra of $H$ and
$j^\#(z)= g_{ab}\r {} a \c {} b +\sum_{\alpha\in{ G\over
H}}\rho^{+\alpha} \chi^{-\alpha}$.
Hence, from the algebraic structure  one  finds that the
\tGH models are indeed TCFT's provided that  $rank\  G =rank \ H$.

\section { BRST cohomology and  Physical states }
Next we  proceed to extract  the space of  physical states of the model.
We take as our definition of a physical state a state
 in  the  \co of  $ Q= Q^{(BRST)} +Q^{G\over H}$, namely, $|phys>\in
H^*(Q)$. In  the case that
the spectral sequence\refmark\BT of the double complex of $Q^{(BRST)}$
and   $Q^{G\over H}$ degenerates at the $E_2$ term, this is the same as taking
one cohomology and then the other.
The extraction of the physical states was worked out
  in detail in refs. [\us,\uss,\usss].  The procedure for the \tGH  is very
 much like the one of the \G models.
 Hence, we summarize here the analysis of the  $A^{(1)}_1$
\G model.
  Expanding the currents $J^a,\  I^a$ and the
$(1,0)$  ghost fields $\rho^a,\ \c { } a$ in modes and inserting them   into
eqn. \mishbGJ\   we obtain the  following BRST charge
$$ Q=\sum_{n=-\infty}^\infty [ g_{ab}\c n a( J_{-n}^b+I_{-n}^b) -\half
f_{abc}\sum_{m=-\infty}^\infty  :\c{-n} a\c{-m}  b  \r{n+m} c: ]\eqn\mishQ$$
where : : denotes normal ordering namely putting  modes with negative
subscripts
to the left of those with positive ones and $\r 0 a$ to the right of $\c 0 a$.
Since  both  $\jt n a $ and $L_n$ are  $Q$ exact
namely  $$ \{Q, \r n a \}  = \jt n  a    \qquad\{Q, G_n \}  =
L_n\eqn\mishQGQQR$$  it follows that
$$L_0\y =  0 \qquad \jt 0 0 \y =  0 \eqn\mishL.$$
 For non-vanishing
eigenvalues of $ L_0$  and $\jt 0 0$  it is easy to see that  $\y$ is
in the image of $Q$ which cannot be true for  a non-trivial
 $|phys>\in H^*(Q)$.

Let us now select a sub-space \F
of the space  of physical states on which   $\r 0 0
=   0$ in addition to   $\jt 0 0 = L_0 = 0$.
 On this sub-space $Q$  which may be written as

$$\eqalign{Q  &= \c 0 0 \jt 0 0  + M \r 0  0 +\hat Q \cr
M &= -\half f_{0bc}\sum_{n\neq 0}  :\c{-n} b\c{n}c :
-\half f_{0bc}  :\c{0} b\c{0}c :,\cr} \eqn\mishhatQ$$  equals
 $\hat Q$.
  We thus start by deducing  $H^*(\hat Q)$ the \co of  $\hat Q$. The states
which correspond  to the latter  are built on a highest weight state
 vacuum  $|J,I>$.  A convenient way
to
handle the $J^a$ and $I^a$ currents is to invoke the following
``bosonization"\refmark\Wak $$\eqalign{J_n^+ &=\b n\cr
J_n^0 &=\sum_m :\b m \g {n-m}:  +{a\over \sqrt{2}}\phi_n\cr
J_n^- &=-\sum_{k,m} :\g m \g k \b {n-m-k}:  -\sqrt{2}a\sum_m\phi_m\g
{n-m} +kn\g n \cr}\eqn\mishwak$$ where $a^2=k+2$.
The fields $\beta$
and $\gamma$ form a bosonic $( 1,0)$ system with $[\g m, \b n
]=\delta_{m+n}$. The modes $\phi_n$ correspond to   the dimension one
operator $i\pa\phi$ and they  obey  $[\p m {} , \p n {} ]=
m\delta_{m+n}$.
In the $I$ sector we use an inverted bosonization
$I_n^0\leftrightarrow  -I_n^0\ , I_n^+\leftrightarrow  I_n^- $,
expressed by operators with tildes $I^-_n =\t b_n$ etc,  and
also take
 $  k\rightarrow -k-4 $. It is easy to realize
that the  highest weight conditions  are obeyed only provided  $\b
0|J,I>=\t\g 0|J,I>  =0$. The normal ordering, however, is with respect
to the usual  $SL(2,R)$ invariant vacuum   $\b
0|J,I>=\t\b 0|J,I>  =0$.
The next step following ref. [\BMP] is to assign a degree  to the various
fields. The idea is  to decompose  $\hat Q$ into terms of different
degrees
  in such a way that there is a nil-potent operator that carries the
lowest degree  which is zero.
An assignment that obeys this requirement is the following
$$\eqalign{ &deg (\c {} {}) =deg (\g  {}) =deg (\t\g  {}) =deg (\p {} +) =1\cr
 & deg (\r {} {})= deg (\b {} ) =deg (\t\b {} ) =deg (\p {}
 -) = -1\cr}\eqn\mishdeg$$

The  spectral sequence  decomposition of $\hat Q$   takes the form

$$\eqalign {\hat Q=& Q^{(0)} +Q^{(1)} +Q^{(2)} + Q^{(3)}\cr
Q^{(0)} =&\sum_n \c {-n} - \b n +2a \sum_{n\ne0} \c {-n} 0 \p n - + \sum_n\c
{-n} + \t\b n
.\cr}\eqn\mishdegQ  $$
where  $\p n {\pm} ={1\over \sqrt{2}}(\p n {}  \pm i\t\p n {})$.
$Q^{(0)}$ is nil-potent on the entire Fock  space.

We proceed now to compute $H^*(Q^{(0)})$,
since, if there is a finite number of degrees for each ghost number,
according to the Lemmas of ref. [\BMP] $H^*(Q^{(0)})$ is isomorphic to the
cohomology of $\hat Q$.
Recall that states in  \F are annihilated by
$$\eqalign{\L_0 = &\hat L_0 +{1\over k+2} [J(J+1) -I(I+1)]\cr
\jt 0 0= & J+I+1+\widehat\jt 0  0 .\cr}\eqn\mishjtot$$
It is easy to check that  $\hat L_0$ and $  \widehat\jt 0  0$ are exact under
$Q^{(0)}$ and therefore both of them
 annihilate the states
in the \co of  $Q^{(0)}$ on  \F. Hence, there are no excitations in
$H^*(Q^{0})$. Moreover, since $L_0 =0$,  states in the latter must have either
$I=J$ or $I=-(J+1)$. Let us now extract the  zero modes contributions  to the
\co . The general structure of these states is
$$|n_\gamma , n_{\t\beta}, n_+ , n_- >=(\g 0)^{n_\gamma} (\t\b
0)^{n_{\t\beta}} (\c 0 +)^{n_+} (\c 0 -)^{n_-} |I,J>\eqn\mishzero$$
 where obviously $n_+, n_-
= 0,1$ and $n_\gamma, n_{\t\beta} $ are non-negative integers.
It is now straightforward to
deduce the Kernel and the Image of  $Q^{(0)}$.
It turns out that  the only possible  state in the relative cohomology
of  $Q^{(0)}$ is $\c 0 + |I,J>$, and from the condition $\jt 0 0 =0$  we find a
state only provided that $I=-J-1$ and then
$$ H^{rel} (Q^{(0)}) = \{ \c 0 + |-(J+1),J> \}. \eqn\mishrelco$$
The passage from the relative \co to the absolute one is  then  given by
$$H^{abs} (Q^{(0)}) \simeq H^{rel} (Q^{(0)}) \oplus\c 0 0  H^{rel} (Q^{(0)})
\eqn\mishabsco$$
in the same way as in ref. [\BMP].
We conclude that the \co of $Q$ on the full Fock space includes states of
arbitrary $J$ with a corresponding $I=-(J+1)$ and with ghost number $G= 0,1$,
where we  have shiftted the definition of the
ghost-number so that the state   $\c 0 + |I,J>$ is at $G=0$.
There is only one state at each ghost number.

 For the general case of a \tGH
the generalization of the  bosonization of eqn. \mishwak\
is performed in terms of the scalars $\phi_i, i=1,...,ran\ G$
  and the commuting $(1,0)$ systems   $(\b {} {\alpha},\g {} {\alpha})\
(\alpha>0)$  $(i\leq j)$.\refmark\uss
Following a similar derivation as that of the \S 2  case one
finds\refmark{\FeFr,\BMP}  the corresponding relative cohomology
 $$ H^{rel} (Q) = \{\prod_{\alpha\in H,\alpha>0}  \c 0
\alpha|\vec J, \vec I>
 ;\   \ \ \vec J + \vec I + 2 \vec \rho_H=0\} .
\eqn\mishrelco$$ where $\vec \rho_H$ is half the sum of the positive roots of
$H$. Since  the scalars $\phi_i$  in the
$J$ sector have a background charge of ${i\over  \sqrt{k+C_G}}\vec
\rho_G\cdot   \pa^2\vec\phi$ while those
in  the $I$ sector have a background charge of
${i\over  \sqrt{k+C_G}}
(2\vec\rho_H-\vec\rho_G)\cdot   \pa^2\vh$, we find that the weight of this
state is
$L_0= {1\over   k+c_G}[\vec J\cdot (\vec J +2\vec \rho_G)-
\vec I \cdot (\vec I +2(2\vec \rho_H-\vec \rho_G))]=0$ for
 $\vec J + \vec I + 2 \vec
\rho_H=0$. This state  corresponds to one of  the ``tachyon" states of the
$W_N$
models based on $G=SU(N)$.\refmark\BLNW\
 The absolute cohomology (without the restriction
$\r 0 i=0$ ) is $$H^{abs} (Q) \simeq H^{rel} (Q)
\oplus\sum_{\{k_1,...,k_l\}}\c 0 {k_1}... \c 0 {k_l} H^{rel} (Q)
\eqn\mishabsco$$ where the sum  is over $\{k_1,...,k_{l}\}$ which are  all
possible subsets of the set  $1,...,rank \ G$. Thus, each state in the relative
cohomology gives rise to $2^{rank\ G}$  states in the absolute cohomology.

\section {  Irreducible representation and the BRST cohomology}
 So far we have
analyzed the cohomology on the whole  Fock Space.
The next step in the extraction of the physical states is to pass
from the \co on the Fock space to  the irreducible representations
of the level $k$   Kac- Moody algebra in the $J$ sector. To simplify the
discussion we  continue to address the  $A_1^{(1)}$ \G model. In general a
representationwith highest weight
 $L$ is reducible iff $ 2L +1=r-(s-1)(k+2) $ where $
r$ and $s$ are integers with either $r,s\ge 1$ or $ r<0, s\leq
0$.\refmark\KK In the \G model with $G= A_1^{(1)}$ we therefore have
reducible representation for $J_{r,s}$ and $I_{r,s}$ where
$$2J_{r,s} +1 =r-(s-1)(k+2)\qquad  2I_{r,s} +1
=r+(s-1)(k+2)\eqn\mishJIrs$$ Note that $J_{r,s} =-I_{-r,s}-1$.
Completely irreducible representations, which have infinitely many
null vectors, appear provided that $k+2={p\over q}$ for $p$ and $q$
positive integers which can be chosen  with no common divisor.  In
this case  $I_{r,s} = I_{r+p, s-q}$ and $J_{r,s} = J_{r+p, s+q}$. It
is, thus, enough to analyze the domain $1\leq s\leq q$,  and we will
choose  $1\leq r\leq p-1$. This choice corresponds to the double line embedding
diagram.\refmark\BF The states corresponding to $r=p$ have a single line
embedding
diagram.
 It was found that \refmark\BF  for the  case of the double line one can
construct the irreducible representation which is contained in $\Frs$, the Fock
space built on $|J_{r,s}>$. This is achieved
 via  the  \co of an
operator $Q_J$  which acts on   $\Frs$    the union of the
Fock spaces that correspond to $J_{r+2lp,s}$ and $J_{-r+2lp,s}$ for every
integer $l$.  It turns out\refmark\BF
 that the relevant information is
encoded in $H^0(\Frs,Q_J)$ and all other levels of the cohomology vanish. The
cohomology is only in the $J$ sector and not in the $I$ sector just as there is
no use of the \co of the Liouville sector in models of $c<1$  matter coupled to
gravity\refmark\BMP. Thus, the space of physical states  of ghost
number $n$ is given by $$H^{(n)}_{rel} [ H^{(0)}(\Frs,Q_J)\times {\cal
F}_I\times {\cal F}^G,Q]\eqn\mishHa$$
where  $ {\cal F}^G$ is the ghosts' Fock space built on the  new vacuum
$|0>_G$.
Since $Q_J$ acts only in the $J$ sector we can rewrite $ H^{(n)}_{rel}$ as
$$H^{(n)}_{rel} [ H^{(0)}(\Frs\times {\cal F}_I\times {\cal
F}^G,Q_J),Q].\eqn\mishHb$$ Moreover, since $\{Q,Q_J\}=0$ one can use
theorems\refmark\BT
  about double cohomologies and write this as isomorphic to
$$H^{(n)} [ H^{(0)}_{rel}(\Frs\times {\cal F}_I\times {\cal
F}^G,Q),Q_J].\eqn\mishHc$$
The theorems\refmark\BT
 apply only provided that each \co separately is
different from zero only for one single degree as  was
shown above.  In fact, we have already calculated $
H^{(0)}_{rel}(\Frs\times {\cal F}_I\times {\cal F}^G,Q) $ since $\Frs$ is the
union of free Fock spaces. Hence the result is that the latter has one state if
the Fock space of $J=-I-1$ is in $\Frs$, and it is empty otherwise.
For each $J_{r,s}$ we get states at $I=-J_{r+2lp,s}-1 $ and
 $I=-J_{-r+2lp,s}-1 $ where their ghost number is equal to the corresponding
degree in the complex of ref. [\BF]
{}.
For each such $J_{r,s}$ there is an infinite set of states with
$I=I_{-r-2lp,s}\ \  G=-2l $ and $I=I_{r-2lp,s}\ \  G= 1-2l$ for every integer
$l$.

Following the same steps for $G=$\S N one finds for each maximal weight
$J$ of $G_k$ a $rank\  G$ dimensional vector of states. This implies also that
there is an $rank\ G-1$ dimensional lattice of states   for each ghost number
and
$J$.
The latter situation  follows
from the $N-1$ dimensional lattice of Fock spaces which are derived by Weyl
reflections as well as shifts by  linear combinations of  roots.\refmark\BMPN
The physical states in the $G=$\S 2 model are characterized by
 $\hat L_0$ and $\hat\jt  0 0 $  as follows

$$\eqalign{ J=J_{r,s},\ \ I= I_{-r-2lp,s},\ \qquad \qquad&
J=J_{r,s},\ \ I= I_{r-2lp,s},\  \cr
 G=-2l\qquad \qquad \qquad \qquad\ \ \ \ \ \ \ \ &G=1-2l\cr\hat L_0=l^2pq
+l(qr-sp) +lp \ \qquad\qquad& \hat L_0=l^2pq -l(qr+sp) +r(s-1)+lp\ \ \cr
 \hat\jt 0  0 =lp\qquad\qquad \qquad \qquad\ \ \ \ & \hat\jt 0   0 =lp-r
\cr}\eqn\mishlevel$$
 vanishes and our ghost vacuum has $J_0=1$.
 For integer $k$ we have $J=0,..,{k\over
2}$.
Let us now examine   the index
interpretation of the torus partition function.\refmark\GK
 We want to check now whether it can be rewritten as a trace
over the space of the physical states.
 One has to  insert the  values of $\hat L_0$ and  $\hat J^0_{(tot)}$ of eqn.
\mishlevel\ into $ Tr[(-)^Gq^{\hat L_0}e^{i\pi\theta \widehat
J^0_{(tot)}} ]$, with $\widehat J^0_{(tot)}$ and $G$ shifted
to the
values defined for  an $SL(2)$ invariant vacuum. Inserting these  values for
every
  $l$, and adding the values of the level and the eigenvalue of $\jt {} 0$ of
the
$|J>$ vacuum,
namely   $\hat L_0 \rightarrow \hat L_o +{J(J+1)\over k+2}$ and
$\widehat\jt {} 0
\rightarrow \widehat\jt {} 0 +J$,
one finds
 $$  Tr[(-)^Gq^{\hat L_0}e^{i\pi\theta \widehat
J^0_{(tot)}} ]= 2iq^{-1\over
4(k+2)}e^{-i\pi {\theta\over 2}}M_{k,J}(\tau,\theta)
.\eqn\mishbMTT$$
where
$$M_{k,j}(\tau,\theta) =\sum_{l=-\infty}^{\infty}
q^{(k+2)(l+{j+\half\over (k+2)})^2}sin\{\pi\theta[(k+2)l +{j+\half}]\}
\eqn\mishbM$$
 is the numerator of the character which corresponds to the
highest weight  state $J$.
We have, thus,  rederived using the BRST  cohomology and the index of above,
the path integral results of ref. [\GK], for the partition function.

   \section { Comparison with
gravitational models}
 The main  motivation to analyse the \tGH models was the idea that they will be
found to be  equivalent to certain gravitational models. More specifically,
we expect a
correspondence between the \A    \tGH models and $W_N$ strings and,
in particular, between  the  case of  $G=SL(2)$  and minimal models coupled to
gravity.
 Let us now check
 whether one  can  map   the topological coset models  into
string models.
In fact, for reasons that will be clarified shortly, it is clear that the
comparison to the gravitational models should be done with the topological
coset models only    after twisting their energy-momentum tensor.
For  $G=$\S N     the
latter is given by
$$T(z)\rightarrow \t T(z) = T(z) +\sum_{i=1}^{N-1}\pa\jt {} i (z)\eqn\mishetT$$
Obviously since $T(z)$ and $\pa\jt {} i (z)$ are BRST exact so is $\t T(z)$.
Thus, the total Virasoro anomaly  is unchanged. However, the contribution of
each
sector to $c$ is modified as follows
$$c_J\rightarrow \t c_J =  c_J-d_GC_Gk \ \  \ \ c_{H^{(I)}}\rightarrow \t
c_{H^{(I)}} =  c_{H^{(I)}}+d_{H^{(I)}}C_{H^{(I)}}(k+C_G+C_{H^{(I)}})
,\eqn\mishtc$$ and the shift in the ghost contribution can be found from a
similar expression or from the  fact that the  sum of the shifts vanishes.
We consider here  for simplicity the case of $G=SL(N,R)$.
The twisted  ghost  sector includes the ghosts of a $W_N$
gravity, namely, a sequence of ghosts with dimensions $(i,1-i)$ for $i=2,...,N$
contributing $\t c_{Wgh}=-2(N-1)[(N+1)^2+N^2] $ to $\t c$.  The rest of the
ghosts are paired with commuting fields of  the same conformal structure
coming from the  $J$ and $I$ sectors. For $N=2$ one finds $\t
c_{Wgh}=-26-2\#_{pairs}$ where there are two pairs in the \G model and one in
the ${SL(2)\over U(1)}$ case.
  The net matter degrees of
freedom have the following  Virasoro anomaly  $c=\t c_J-\half[\t
c_{(gh)}-\t c_{Wgh}]=(N-1)[(2N^2 +2N +1) -N(N+1)(t+{1\over
t})]$ which is exactly that of a $(p,q)$ minimal $W_N$ matter
sector\refmark\FaLu provided  $t\equiv k+N={p\over q}$.
This was explicitly verified by analyzing the dimensions and contributions
to $\t c$ of the various free fields in the $J$ sector\refmark\uss.
The expression for $c$ reduces to that of the ${p,q}$ minimal model ( plus 2)
for $N=2$.
Before twisting
 the contribution of the set of all  $\p{} i$ to $c$ is $c_\phi = (N-1)
-{12\sum_{i,j}g^{ij}\over k+N}= -(N-1)[{(N^2+N)\over t} -1]$.
Due to the twisting the  $(\b {} {(ij)},\g {} {(ij)})$\refmark\nota systems
acquire dimensions of $(i-j, j-i+1)$ and thus there are $N-1 $  systems of
dimension $(0,1)$,   $N-2$ pairs of fields of dimension  $(-1,2)$   up to one
pair of dimensions $(2-N,N-1)$ .
The $\phi$ central charge is modified to
$$\t c_\phi
=(N-1)[(2N^2 +2N +1) -N(N+1)(t+{1\over t})]\eqn\mishcm$$ which is
identical to the net matter contribution to $c$ given above and hence
the $\phi^i$ fields are in fact those of the $W_N$ model. A further
indication of the  latter equivalence  is
 the dimensions of the $\phi$ fields which correspond to the maximal
weights $\lambda^j=\sum_k g^{jk}[(r_k-1)-t(s_k-1)]$ which after the twisting
are
$$\Delta_{r_1,...,r_{N-1},s_1,...,s_{N-1}} = { 12\sum_{i,j} g^{ij}
(ps_i-qr_i)(ps_j-qr_j)-N(N^2-1)(p-q)^2\over 24 pq}\eqn\mishDelta$$ as in the
$W_N$ minimal models.\refmark\FaLu\  If one parametrizes the $I$ sector in the
same way as  the $J$ sector, then  clearly  the modified dimensions of the
$(\t\b {} {(ij)},\t\g {} {(ij)})$ fields are the same as those without tilde.
{}From the point of view of their contribution to $\t c$   the $\t \p {} i$
fields  are then identical to those of $W_N$ gravity. This is achieved by
replacing $t$ with $-t$ in $\t c_\phi$ defined above.
 With respect to the untwisted $T$ the ghosts $(\r {} a,\c {} a)$ are
all of dimension $(1,0)$. The   ghost part of $\t T(z)$ has the form
$$ \eqalign{\t T^{(gh)}(z) =&g_{ab}:\r {} a \pa \c {} b:\cr
+&\sum_{1\leq i\leq j\leq N-1} (j-i+1)\pa[:\r {} {-(ij)}(z) \c {} {+(ij)}(z):-
:\r {} {+(ij)}(z) \c {} {-(ij)}(z):].\cr}\eqn\mishTgh$$ It is thus obvious that
the members of the Cartan sub-algebra $\r {} i, \c {} i$ remain $(1,0)$ fields.
On the other hand the pair $(\c {} {+(ij)}, \r {} {-(ij)})$  carries now
dimensions $(i-j-1, 2+j-i)$ and  $(\c {} {-(ij)}, \r {} {+(ij)})$ carry
dimensions $(j-i+1, i-j)$. Altogether one finds for the $(\chi, \rho)$ ghosts
$N-1$ pairs of fields of dimension $(0,1)$ coming from the Cartan sub-algebra,
$N-1$ pairs of dimension $(-1,2)$
$N-2$ pairs of dimension $(-2, 3)$ up to a  pair  of  dimension $(1-N,N)$, and
similarly $N-1$ pairs of dimension $(1,0)$ up to one pair of $(N-1,2-N)$.
It is now clear that  when the dust settles the \G model  of  \S N at level
$k={p\over q}- N$ has the field content of a minimal  $W_N$ $(p,q)$ model
coupled to  $W_N$ gravity  plus pairs of ``topological sectors" namely pairs of
commuting and anti-commuting $(i,1-i)$ systems for $i=1,...,N-1$.

 Next we want to compare the partition function of the $(p,q)$
model to that of \G for  $G=SL(2)$ and  $k= {p\over q}-2$. Comparing eqn.
\mishbMTT\ to the
numerator of the character of the minimal model it is clear that correspondence
might be achieved only provided one takes $\tau=-\half\theta$.
Recall that in the topological coset models we integrate in the path-integral
only over $\theta$ ( and not over $\tau$) and the result is $\tau$
independent.\refmark\GK
 In this case  the numerator of the
character in the minimal model which is proportional to $Tr[(-1)^G q^{\hat
L_0}]$ is mapped into  $Tr[(-1)^G u^{\hat L_0 -\widehat \jt 0 0 }]$ in the \G
model. The integration over the moduli parameter of the torus is therefore
replaced by the  integration over the moduli of flat gauge connection.
  We thus
need to compare the number of states at a given level and ghost number in the
minimal models with the corresponding numbers at the same ghost number and
``twisted level". From eqn. \mishlevel\  we read
 $$\eqalign{
 I= I_{-r-2lp,s}\ \ G=-2l \qquad&\hat L_0 -\widehat\jt 0 0=l^2pq
+l(qr-sp) \cr
 I= I_{r-2lp,s}\ \ G=1-2l \qquad&\hat L_0 -\widehat\jt 0 0=l^2pq
-l(qr+sp)  +rs\cr
}.\eqn\mishdL$$
In the minimal models we have states built on vacua
labeled  by the pair $r,s$ with $1\leq r\leq p-1$ and $1\leq s\leq q-1$
with $ps>qr$ which have dimension $h_{r,s} = {(qr-sp)^2-(p-q)^2\over 4pq}$.
The levels of the excitations  are $\hat L_0=\Delta-h_{r,s}$.
  For $G=2l+1$ one has $\Delta=A(l)={[(2pql+qr+sp)^2-(q-p)^2]\over 4pq}$ and
for $G=2l$ $\Delta=B(l)={[(2pql-qr+sp)^2-(q-p)^2]\over 4pq}$.
\refmark{\BerK,\LZ}
Hence,
the  the contribution of the various levels to the partition function
 are identical to those of $\hat L_0 -\widehat\jt 0 0$ in  eqn. \mishdL\ for
 the
same ghost numbers and the respective vacua satisfy $J=\sqrt{p\over 2 q}p_m$
and
$I=-\sqrt{p\over 2 q}p_L$ where $p_m$ and $p_L$ are the matter and Liouville
momenta respectively. It is thus clear that for a given $r,s$ we get the same
number of states with the
same ghost number parity in the two models and thus the two partition functions
on the torus are in fact identical.
  To obtain
the partition function of the $(p,q)$ models coupled to 2d gravity we have
restricted $r$ and $s$ as follows $1\leq s\leq q-1$ and  $1\leq r\leq p-1$. It
is interesting to note
that we could include in the sum over $r$ and $s$ which appears in the
partition function also the terms with $r=p$. Those terms arise from the
states  which have a single line as their embedding diagram. The $r=p$ terms
cancel between themselves and do not change the result for the partition
function.

\section { discussion}
The description of gravitational ($W_N$ gravity) models via topological
coset models is far from being complete.
A comparison of the correlators  of the latter with those of other formuations
of 2d gravity ($W_N$ gravity) is still missing. Another challenge is to
extract information on the behaviour of the models on higher genus Riemann
surfaces.  The differneces between models which correspond to various $H$ for a
given $G$  are only partially understood. Possible flows between these models
and in general ``renormalization" flows
in the space of TCFT's are under current study.
 The precise relations between the present formulation and the known
results on   chiral rings of $N=2$\refmark\Gepner theories deserve further
investigation.
 \refout
\end
%****************************************************************************
\bye